\documentclass[journal]{IEEEtran}
\usepackage{graphicx}
\usepackage{subfigure}

\begin{document}
\title{Development of CMOS monolithic pixel sensors with
in-pixel correlated double sampling and fast readout for the ILC}
%
%

\author{Marco Battaglia,
        Jean-Marie Bussat,
        Devis Contarato,
        Peter Denes,
        Piero Giubilato,
        Lindsay E. Glesener
\thanks{Manuscript received November 23, 2007.
This work was supported by the Director, Office of Science, of the
U.S. Department of Energy under Contract No.DE-AC02-05CH11231.}
\thanks{M. Battaglia is with the Department of Physics, University of 
California and the Lawrence Berkeley National Laboratory, Berkeley, CA 94720, 
USA (telephone: 510-486-7029, e-mail: mbattaglia@lbl.gov).}%
\thanks{J.M. Bussat was with the Lawrence Berkeley National Laboratory, 
Berkeley, CA 94720, USA.}%
\thanks{D. Contarato is with the Lawrence Berkeley National Laboratory, 
Berkeley, CA 94720, USA}%
\thanks{P. Denes is with the Lawrence Berkeley National Laboratory, 
Berkeley, CA 94720, USA}%
\thanks{P. Giubilato is with the Istituto Nazionale Fisica Nucleare, Sezione 
di Padova, Italy and a visitor at the Lawrence Berkeley National Laboratory, 
Berkeley, CA 94720, USA}%
\thanks{L.E. Glesener is with the Lawrence Berkeley National Laboratory, 
Berkeley, CA 94720, USA}%
}

\maketitle
\pagestyle{empty}
\thispagestyle{empty}

\begin{abstract}
This paper presents the design and results of detailed
tests of a CMOS active pixel chip for charged particle
detection with in-pixel charge storage for correlated
double sampling and readout in rolling shutter mode at
frequencies up to 25~MHz. This detector is developed in 
the framework of R\&D for the Vertex Tracker for the 
International Linear Collider.

\end{abstract}


\section{Introduction}

\IEEEPARstart{T}{he} Vertex Tracker for the International Linear Collider (ILC) 
has requirements in terms of position resolution and material budget 
that largely surpass those of the detectors at LEP, SLC and LHC. 
The single point resolution needs to be $\le$~3~$\mu$m and the detector has be read-out 
fast enough that the machine-induced background does not adversely affect the track 
pattern recognition and reconstruction accuracy. Detailed simulation of incoherent pair 
production in the strong field of the colliding beams, the dominant background source, 
predicts an hit density of 5~hits~bunch~crossing$^{-1}$~cm$^{-2}$ on the innermost detector 
layer, located at a radius of 1.5~cm, for a solenoidal field of 4~T. 
The requirement of an occupancy $\le$0.1~\%, corresponds to a maximum 80 bunch crossings 
which can be integrated  by the detector in a readout cycle, i.e.\ a readout frequency of 
25~MHz for a detector with 512 pixel long columns.
Finally, power dissipation must be kept small enough so that cooling can be achieved 
by airflow, without requiring those active cooling systems which largely contribute to 
the material budget of the vertex detectors installed in the LHC experiments. 
Tests performed on a carbon composite prototype ladder, equipped with 50~$\mu$m-thin CMOS 
pixel sensors, have shown that an airflow of 2~m~s$^{-1}$ can remove 
$\simeq$ 80~mW~cm$^{-2}$. Now, assuming a pixel column made of 512 pixels, this 
corresponds to a maximum allowable power dissipation of $\simeq$0.5~mW~column$^{-1}$.

An attractive sensor architecture for the ILC Vertex Tracker sensor is a
pixel of $\simeq$20$\times$20~$\mu$m$^2$ readout at 25-50~MHz during 
the long ILC bunch train. Signals are digitised at the end of the column with 
enough accuracy to allow charge interpolation to optimise the spatial resolution.
Digitising at the required speed and within the maximum tolerable power dissipation 
poses major design challenge. The study of data collected with a CMOS pixel test 
chip, with 10$\times$10~$\mu$m$^2$, 20$\times$20~$\mu$m$^2$ and 
40$\times$40~$\mu$m$^2$ pixels~\cite{snic} shows that a 5-bit ADC accuracy is 
sufficient, provided that pixel pedestal levels are subtracted before digitisation. 

\section{LDRD-2: A pixel chip with in-pixel CDS}

We have designed a CMOS monolithic pixel chip with in-pixel correlated 
double sampling (CDS) and tested it for readout speeds up to 25~MHz. 
The chip consists of a matrix of 96$\times$96 pixels arrayed on a 20~$\mu$m 
pitch. Each pixel has two 5$\times$5~$\mu$m$^2$ PIP capacitors, corresponding 
to a capacitance of $\simeq$20~fF. These capacitors are used for the storage of 
the pixel reset and signal levels. The net pixel signal is obtained by 
subtracting the reset from the signal level. In the current CDS implementation, 
this subtraction is performed on-line, allowing for a detailed performance study.
Pixels are readout in rolling shutter mode, which ensures a constant integration 
time across the pixel matrix. The pixel array is divided in two 48$\times$96 pixel 
sections, which are readout in parallel. Different pixel designs, including 
diode sizes of 3~$\mu$m and 5~$\mu$m, have been implemented. The detector has 
been fabricated in an AMS 0.35~$\mu$m 4-metal, 2-poly CMOS-OPTO process, which 
provides an epitaxial layer with a nominal thickness of 14~$\mu$m.

The readout sequence is as follows. On a reset signal on the pixel $i$, 
the pixel reset level is stored and the charge in integrated on the diode, while the 
reset level is stored for the pixel $i+1$. After the integration time which corresponds
to the scan of a section, i.e.\ $N_p/f$, where $N_p$ is the number of pixels in a readout 
section and $f$ the readout frequency, the signal level is stored, the pixel is reset and 
the cycle restarted. The reset and signal levels are readout serially. The highest 
frequency tested is 25~MHz, corresponding to an integration time of 184~$\mu$s. 
In order to minimise the power dissipation of the pixel cell, the source follower is 
switched on only in the short time elapsing between the write signal level of one 
event and the write reset level of the next event.

The detector is readout through a custom FPGA-driven acquisition board.
A set of 14~bits, 40~MSample/s ADCs reads the chip outputs, while an array of 
digital buffers drives all the required clocks and synchronisation signals. 
The FPGA has been programmed to generate the clocks pattern and collect the 
sampled data from the ADCs. A 32~bits wide bus connects the FPGA to a
digital acquisition board installed on a control PC. Data is processed on-line
by a LabView-based program, which performs the subtraction of the reset level 
from the pixel level and computed the pixel noise and residual pedestal.  

The detector has been tested in the lab using both a 1.2~mCi $^{55}$Fe collimated 
source and a 850~nm IR pulsed laser. 
The detector performance has been studied as a function of the readout 
frequency, from 1.25~MHz up to 25~MHz. First the pixel noise has been measured 
for operation at room temperature. No significant degradation of the noise of 
the pixel matrix is observed up to the highest frequency. The measured noise of 
$\simeq$45~ENC is due in part to the readout electronics noise. The pixel noise 
has also been studied as a function  of the operating temperature (from -10~$^o$C to 
+30~$^o$~C) for a readout frequency of 1.25~MHz. 

\begin{figure}[!h]
\centering
\includegraphics[width=3.0in]{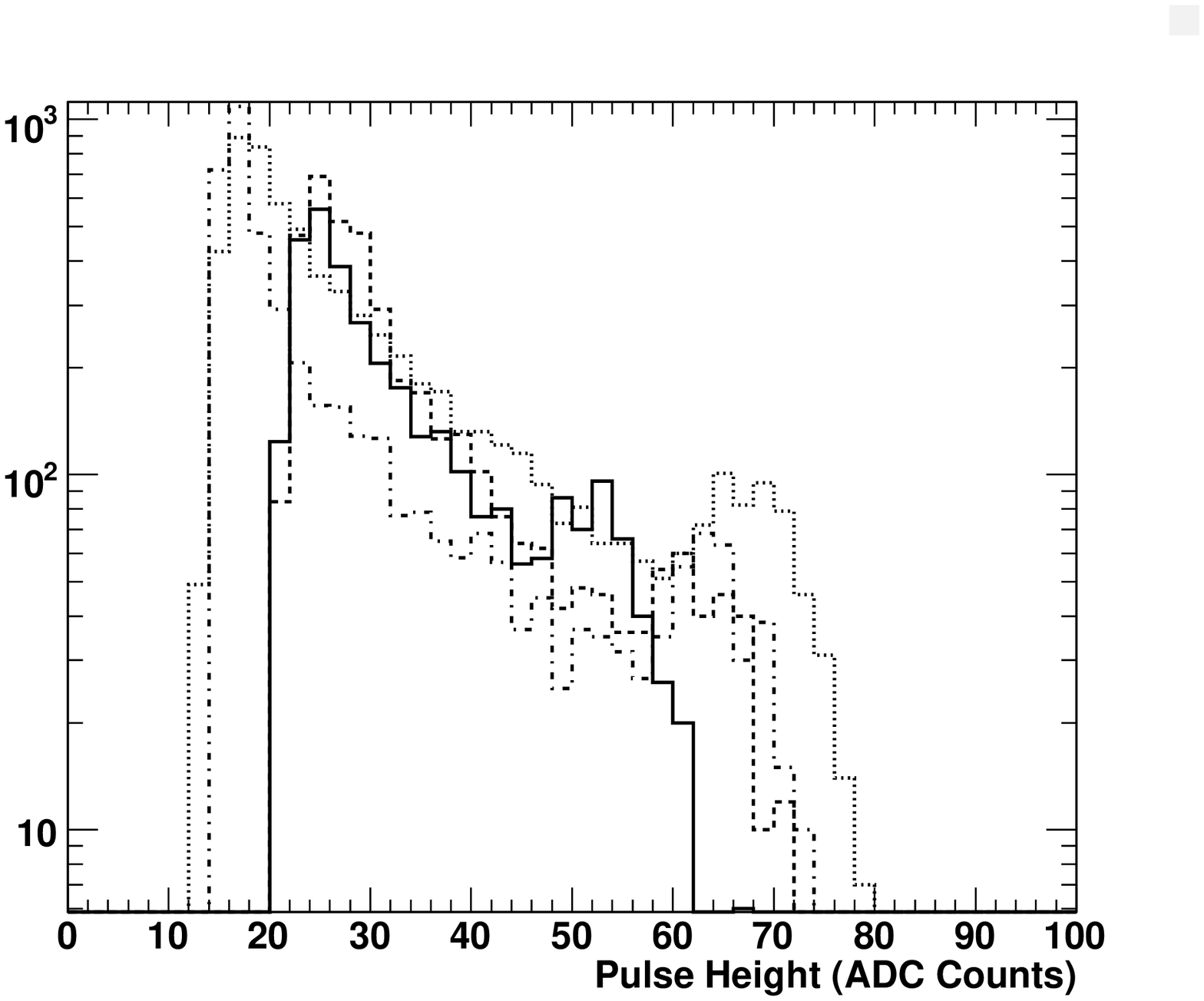}
\caption{Response of the LDRD-2 chip to 5.9 X-rays from $^{55}Fe$ for readout 
frequencies of 1~MHz (dotted), 6.25~MHz (dash dotted), 12.5~MHz (dashed) and 
25~MHz (continuous).}
\label{fig:55fe}
\end{figure}  

The pixel calibration has been studied using the 5.9~keV X-rays from a $^{55}$Fe 
source for various readout frequencies. Charge generated by X-rays which convert in 
the shallow depletion region near the pixel diode is fully collected, resulting in a
pulse height peak corresponding to the full X-ray energy, or 1640 electrons. Pulse 
height spectra recorded on a single pixel for selected clusters are shown in 
Figure~\ref{fig:55fe}, for different readout frequencies.

The response to high momentum particles has been studied
in beam tests. We used both the 1.3~GeV electron beam from the LBNL Advanced Light 
Source (ALS) booster and the 120~GeV secondary proton beam at the Meson Test Beam 
Facility (MTBF) at Fermilab, as part of the T966 beam test experiment. 
Data are converted into the {\tt lcio} format, which is the persistency format 
adopted by the ILC studies. The data analysis is performed offline by a dedicated 
set of processors developed in the {\tt Marlin} C++ framework.
Events are first scanned for noisy pixels. The noise and pedestal values 
computed on-line are updated, using the algorithm in ~\cite{chabaud}, to 
follow possible variations in the course of a data taking run.
\begin{figure}[!h]
\centering
\includegraphics[width=3.5in]{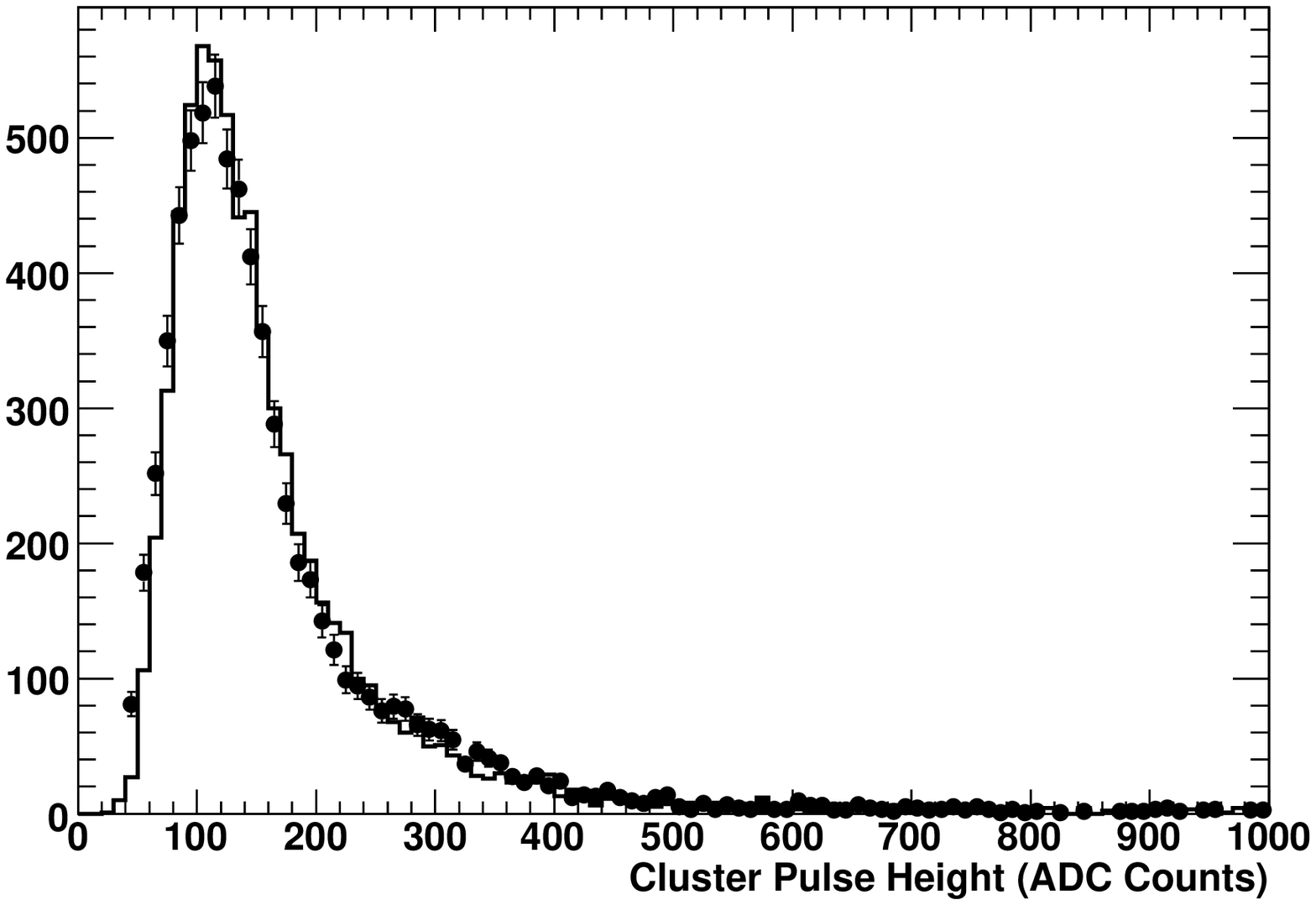}

\includegraphics[width=3.5in]{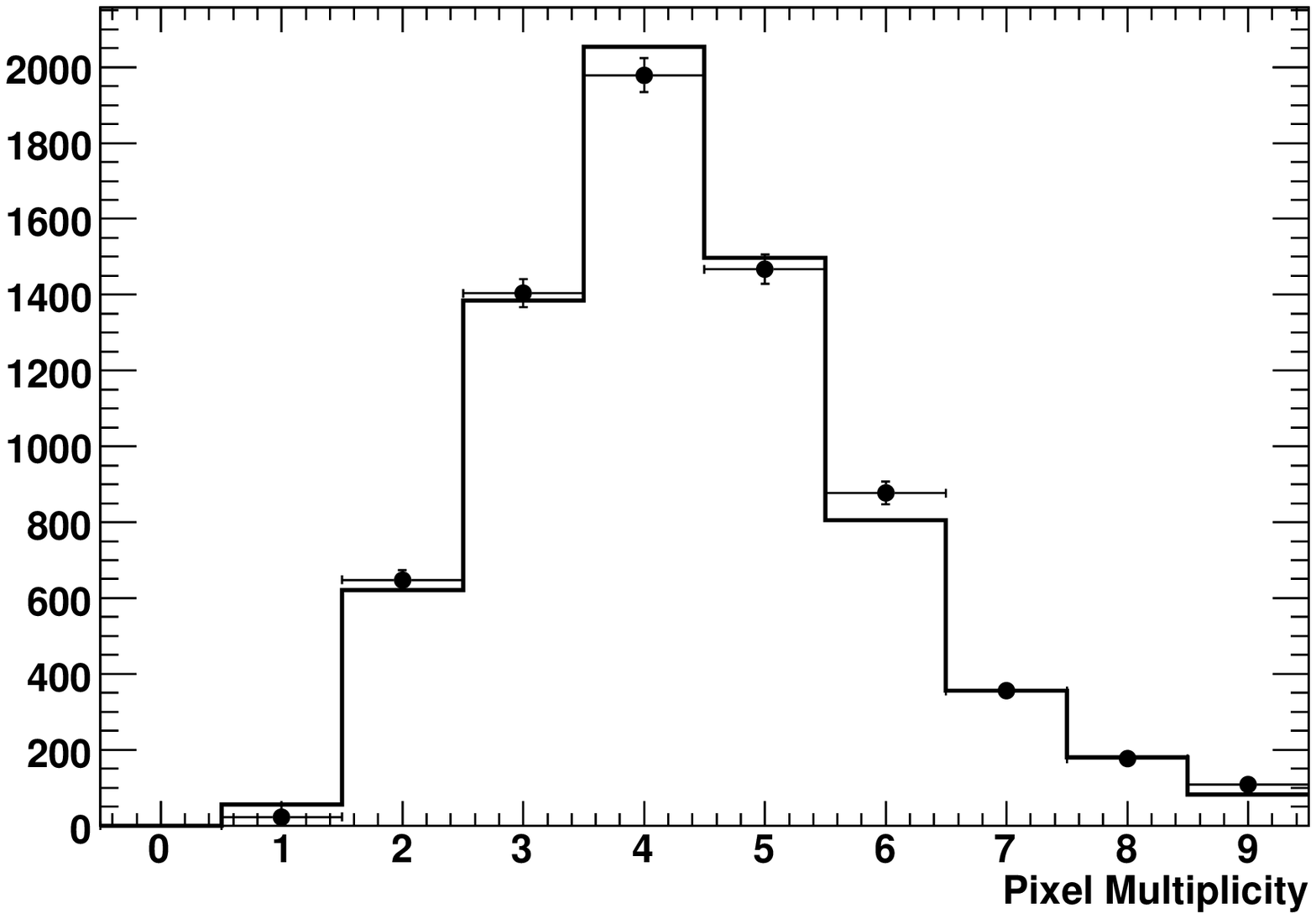}
\caption{Response of the LDRD-2 chip to 1.35~GeV $e^-$s: Data collected at the ALS
are compared to predictions from a simulation based on {\sc Geant-4} a dedicated 
sensor simulation in {\tt Marlin} for the cluster pulse height (above) and pixel 
multiplicity in the cluster (below).}
\label{fig:sim}
\end{figure}  
Cluster search is performed next. Each event is scanned for pixels with pulse 
height values over a signal-to-noise (S/N) threshold of 5, these are designated 
as cluster `seeds'.  Seeds are then sorted according to their pulse heights and 
the surrounding, neighbouring pixels are tested for addition to the cluster. 
The neighbour search is performed on a 7$\times$7 matrix surrounding the pixel 
seed and the neighbour threshold is set at 2.5, in units of the pixel noise.     
Clusters are not allowed to overlap, i.e. pixels already associated to one 
cluster are not considered for populating another cluster around a different 
seed. Finally, we require that clusters are not discontinuous, i.e. pixels 
associated to a cluster cannot be interleaved by any pixel below the 
neighbour threshold. The pixel response is simulated with a dedicated digitisation 
processor in {\tt Marlin}~\cite{pixelsim}. The charge collection process is 
described starting from ionisation points generated along the particle trajectory 
using {\tt Geant~4}~\cite{Agostinelli:2002hh}, by modelling the diffusion of charge 
carriers, originating in the epitaxial layer, to the collection diode. Simulated 
data are then processed through the cluster reconstruction stage, using the same 
processor as beam test data. A comparison of the real and simulated cluster pulse 
height and pixel multiplicity obtained, after simulation tuning, is shown in 
Figure~\ref{fig:sim}

\begin{figure*}[!t]
\centerline{\includegraphics[width=5.95in]{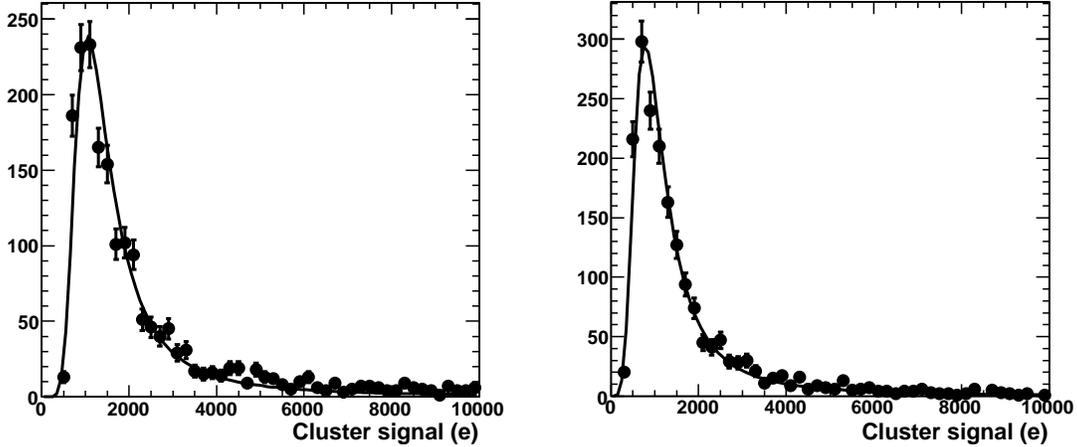}}%
\caption{Response to 120~GeV protons: Cluster pulse height 
distributions obtained for pixels with 5~$\mu$m (left) and 
3~$\mu$m (right) diode. The continuous lines show the 
interpolated Landau functions.}
\label{fig:landau}
\end{figure*}

\begin{figure}[!h]
\centering
\includegraphics[width=3.75in]{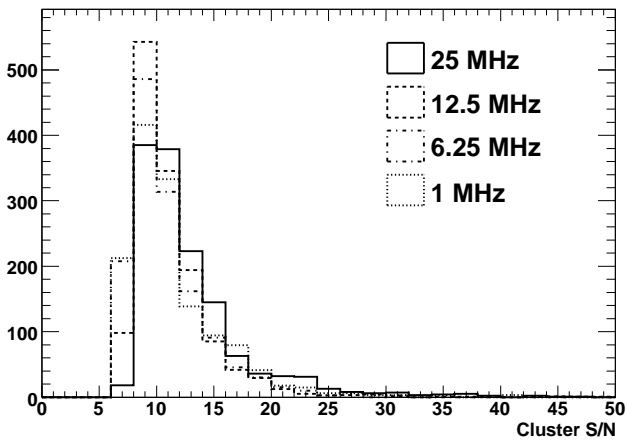}
\caption{Response of the LDRD-2 chip for various 
readout frequencies: cluster signal-to-noise distributions 
for 1.35~GeV $e^-$s for readout 1~MHz (dotted), 6.25~MHz (dash dotted), 
12.5~MHz (dashed) and 25~MHz (continuous).}
\label{fig:sn}
\end{figure}

The most probable cluster pulse height for 120~GeV protons has been measured to 
be (1112$\pm$17)~$e^-$s for 5~$\mu$m and (830$\pm$13)~$e^-$s for 3~$\mu$m collecting 
diodes as shown in Figure~\ref{fig:landau}. The most probable pixel multiplicity 
scales from 2.7 to 4.3, respectively. An average signal-to-noise ratio of 12 to 
13 has been measured for 1~MHz and 25~MHz readout, respectively (see Figure~\ref{fig:sn}). 
The signal-to-noise performance is limited in part by the noise of the read-out 
board.
Finally, the response to low momentum electrons has been studied. The response
of the detector to low energy electrons is important since particles from pair 
background have energies typically below 100~MeV. Data has been collected at the 
ALS with an Al beam scraper placed few meters upstream of the LDRD-2 detector. 
A large fraction of the registered hits are due to low energy electrons, originating 
from the interaction of the primary beam with the Al scraper, or to tertiaries,
from photon conversions. These hits are characterised by large, asymmetric clusters, 
which suggest that low-energy background hits could be identified and rejected, 
based on the observed cluster shape.

\section{Conclusion}

The LDRD-2 pixel cell, with small pitch and in-pixel charge storage for correlated 
double sampling, offers a viable architecture for a pixel chip with analog, fast 
readout, with potentially able to match the requirements for an ILC Vertex Tracker.
Test chips have been characterised using $^{55}$Fe, an IR laser an particle beams.
The chip performance is found to be stable up to the highest tested readout frequency 
of 25~MHz, corresponding to an integration time of 184~$\mu$s. The pixel cell is the base 
of a chip of third-generation, which further addresses the ILC requirements. 
The LDRD-3 chip consists of a matrix of 96$\times$96 pixels on a 20~$\mu$m pitch, 
with the same pixel cell as the LDRD-2. In addition, the chip features a column parallel 
readout at frequency up to 50~MHz and the digitisation performed on-chip, at the end of 
each column, by a row of successive approximation, fully differential ADCs featuring 
low power dissipation. Each ADC has a size of 20~$\mu$m$\times$1~mm, which matches the 
pixel pitch. The digitisation of the 96 rows is performed in 1.9~$\mu$s. The chip has been 
produced in AMS 0.35~$\mu$m CMOS-OPTO process. Chips have been received back from the 
foundry in October 2007 and are currently being characterised.





\end{document}